\documentstyle[psfig,twoside,fleqn,espcrc2]{article}

\newcommand{\AmS}{{\protect\the\textfont2
  A\kern-.1667em\lower.5ex\hbox{M}\kern-.125emS}}

\title{Next-to-Leading Order QCD corrections \\to Three-Jet Rates
       with Massive  Quarks\thanks{Research supported by BMBF,
         contract 057AC9EP(1).}}

\author{W. Bernreuther\thanks{Talk given at the workshop on Standard Model
Physics,
 San Miniato, April 1997.},
         A. Brandenburg\thanks{supported by Deutsche
Forschungsgemeinschaft.}, and P. Uwer\\ \vspace{.5cm}
         Institut f. Theoretische Physik der RWTH Aachen, D-52056 Aachen,
Germany \\ \vskip -5cm hep-ph/9709281 \hfill PITHA 97/35}

\begin{document}

\begin{abstract}
\vskip 4.5cm
The reaction $e^+e^-$ annihilation into three jets was recently computed for
massive quarks at next-to-leading order in perturbative QCD. We discuss some results of
our calculation for $b$ jets produced at the $Z$ resonance. 
\end{abstract}

\maketitle

\section{INTROCUCTION}

Huge samples of $Z$ decays into jets of hadrons have been collected both at LEP and at SLC. 
These data allow for numerous precision tests of  the theories of electroweak and strong interactions.
In particular, large numbers of  jet events involving $b$ quarks can be  isolated using vertex detectors.
For detailed investigations of $b$ jets, aiming at a precision at the few percent level,
one must take into account in theoretical predictions
that the $b$ quark mass is non-zero. In fact $b$ quark mass effects in jets produced at the $Z$ resonance 
have already been seen indirectly in the so-called ``flavour-independence tests" of the 
strong coupling. Such tests have been performed both at LEP \cite{LEP} and at SLC \cite{NLC}.

Three groups have recently reported \cite{Ac,Val,Mil} the calculation of the next-to-leading order (NLO)
QCD corrections for $e^+ e^-\to 3$ jets for massive quarks. This extends the well-known  NLO 3-jet results
for massless quarks obtained in the early eighties \cite{ERT,FSKS,Ver}.  (For the NLO
 2-jet cross section  and  leading order (LO) 
results for  three, four, and five jet rates involving massive quarks see \cite{JLZ} 
and \cite{Joffe,BMM}, respectively.)
In this talk we discuss some 
results of our NLO computation of $\sigma_{\rm{3jet}}$ and apply it to $b$ jets produced at the $Z$ resonance.

\section{THE NLO CROSS SECTION}

The calculation
of the $e^+ e^-$ annihilation cross section, $\sigma_{NLO}^{3}$,  
 into three jets involving a massive
quark antiquark pair
to order $\alpha_s^2$ consists of two parts:
First, the computation of the amplitude of the partonic reaction
$e^+e^-\rightarrow\gamma^\ast, Z^\ast\rightarrow Q \bar Q  g$
at leading and next-to-leading order in the  QCD coupling.  
Here $Q$ denotes a massive quark
and $g$ a gluon. We have calculated the complete decay amplitude and decay
distribution structure for this reaction. This allows for
predictions including
oriented three-jet events. The differential cross section involves the
so-called hadronic
tensor which contains five parity-conserving and four parity-violating
Lorentz structures.
Second, the leading order matrix elements of the
four-parton production processes
$ e^+e^- \rightarrow Z^\ast, \gamma^\ast \rightarrow
ggQ\bar{Q},Q\bar{Q}q\bar{q},Q\bar{Q}Q\bar{Q}$ are needed.
Here $q$ denote light quarks which are taken to be massless.
\par
The infrared (IR) and ultraviolet (UV) singularities, which are encountered
in the computation of the one-loop integrals,
are treated within the framework
of dimensional regularization in $D=4-2\epsilon$ space-time dimensions.
We remove the UV singularities  
by the standard $\overline{\rm MS}$ renormalization.
We have converted from the outset the on-shell mass of the heavy quark $Q$
into the corresponding running $\overline{\rm MS}$ mass. It is known that far
from threshold one thereby absorbs some large logarithms into the running mass.

An essential aspect of any NLO computation of jet rates is to show 
that the IR singularities of the virtual corrections are cancelled by the
singularities resulting from 
phase space integration of the squared tree amplitudes
for the production of four partons.
Different methods to perform  this cancellation have been developed
(see \cite{GG,FKS,CS} and references therein).
We use the so-called phase space slicing method elaborated in
\cite{GG}. The basic idea is to ``slice'' the phase space of the
four parton final state by introducing an unphysical parton
resolution parameter $s_{min}\ll sy_{cut}$, where $y_{cut}$ is
the jet resolution parameter. The parameter  $s_{min}$ splits the 
phase space into
a region where all four partons are ``resolved''
and a region where at least
one parton remains unresolved.
For massless partons, the resolved region may be conveniently
defined by the requirement that all invariants
$s_{ij}=(k_i+k_j)^2$ constructed from the parton momenta $k_i$ are
larger than the parameter
$s_{min}$. We have modified
this definition slightly to account for masses.
\par
In the unresolved region soft and collinear
divergences reside, which have to be isolated explicitly to cancel
the singularities of the virtual corrections.
This is considerably simplified due to collinear and soft
factorizations of the matrix elements
which hold in the limit $s_{min}\to 0$.
(In the presence of massive quarks, the structure of
collinear and soft poles
is completely different as compared to the massless case.)
After having cancelled these IR poles  
against the IR poles of the
one-loop integrals entering the virtual corrections,
one is  left with a completely regular 
differential three-parton cross
section which depends on $s_{min}$.
\par
The contribution to $\sigma_{NLO}^3$ of 
the ``resolved'' part of the four-parton cross section is finite
and may be evaluated in $D=4$ dimensions, which is
of great practical importance.
It also  depends on $s_{min}$ and is  most conveniently obtained 
by a numerical integration.
Since the parameter $s_{min}$ is
completely arbitrary, the sum of all contributions 
to $\sigma^3_{NLO}(y_{cut})$ must not depend on $s_{min}$.
In the soft and collinear approximations 
one neglects terms which vanish
as $s_{min}\to 0$. This limit can  be carried out numerically.
Since the individual
contributions depend logarithmically
on $s_{min}$, it is a nontrivial test of the
calculation to demonstrate that
$\sigma_{NLO}^3$ becomes independent of
$s_{min}$ for small values
of this parameter \cite{Ac}.
\par
The  three jet cross section depends on the
experimental jet definition.
We consider here the JADE \cite{Jade} and Durham \cite{Dur}
clustering algorithms, although other schemes \cite{BKSS} can also be
easily implemented. We have checked that 
we recover the result of \cite{KN} in the massless limit.
\par
Let us now discuss some results for
the cross section $\sigma_{NLO}^{3,b}$ for $b$ quarks.
An important point to notice  is that we are concerned here
with the computation of $tagged$ cross sections -- whereas  the
massless NLO results \cite{ERT,FSKS,Ver} apply to summing over all  quark 
flavours in the final state. In the computation of tagged cross sections
one encounters mass singularities in the real corrections -- that are regulated by keeping the quark mass
non-zero -- which find no counterpart in the virtual corrections
against which they can cancel. In the case at hand, to order $\alpha_s^2$, there is a
contribution to $\sigma_{NLO}^{3,b}$ from the diagrams
$e^+e^-\to q\bar{q}g^*\to q\bar{q}b\bar{b}$. 
This contribution contains large logarithms 
involving $m_b$. 
(They result from the region where the invariant mass of the virtual gluon 
becomes small.) One may simply keep these logarithms, but that is
not satisfactory, in particular if $\sqrt s \gg m_b$. Eventually, one has
to factorize these logarithms into a fragmentation function for a gluon 
into a B hadron. A detailed discussion of the single $b$-tag cross section
$\sigma_{NLO}^{3,b}$ will be given elsewhere \cite{Ac2}.

Here we shall consider instead the following  
``double $b$-tag"
three-jet cross section: we require that at least two 
of the jets that remain after the 
clustering procedure contain a $b$ or $\bar{b}$ quark.  
The cross section $\sigma_{NLO}^{3,b}$, where the contributions from the above
diagrams are included, remains infrared-safe in the limit $m_b\to 0$.

As mentioned above we have expressed $\sigma_{NLO}^{3,b}$ in terms of the $b$ quark mass parameter  $m_b(\mu)$
defined in the  $\overline{\rm MS}$ scheme at a scale $\mu$.
The asymptotic freedom property of QCD predicts that
this mass parameter decreases when being evaluated at a higher scale.
(A number of low energy determinations of the $b$ quark mass have
been made; see for instance \cite{Nar,Neubert,Lattice} and references therein.)
With $m_b(m_b)$ = 4.36 GeV \cite{Neubert} 
and $\alpha_s(m_Z)$ = 0.118 \cite{PDG}
as an input and employing the
standard renormalization group evolution of the coupling and
the quark masses, we use the value
$m_b(\mu=m_Z)$ = 3 GeV. 
These values for $\alpha_s$ and $m_b$ are used in Figs. 1a,b where we plot 
$\sigma_{NLO}^{3,b}$ as a function of $y_{cut}$
together with the LO result at the $Z$ peak, both for the JADE and the Durham
algorithm. 
\unitlength 1cm
\begin{picture}(7.5,5)
\put(0,-1.2){\psfig{figure=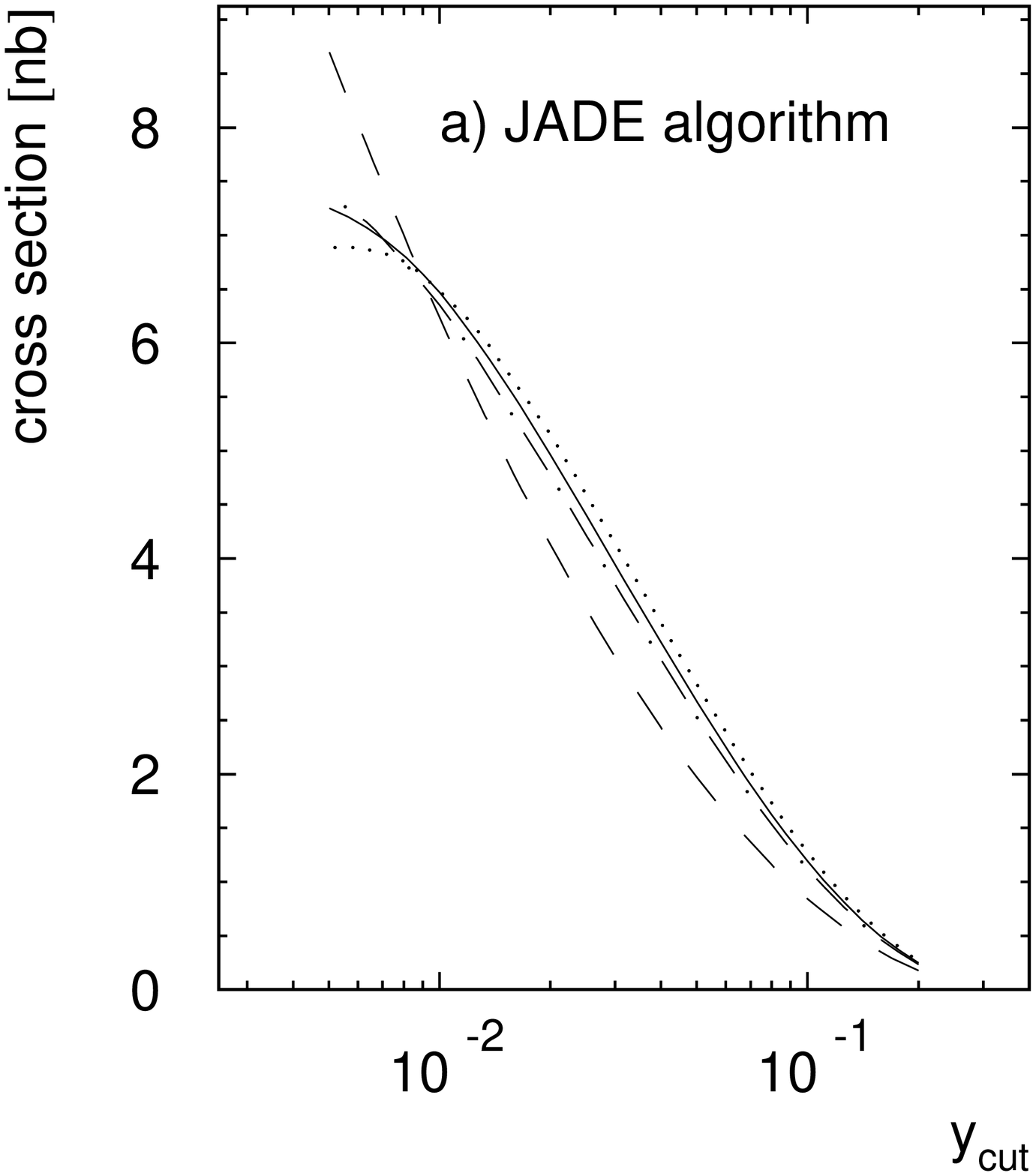,height=6.05cm,width=5.5cm}}
\put(0,-1.2){\begin{minipage}[t]{15cm}
\baselineskip14pt
\small {\bf{Figure 1:}}
The cross section $\sigma^{3,b}$ 
as a function of $y_{cut}$ for the JADE and Durham algorithms,
respectively.
The dashed line is the LO result.
The NLO results are for renormalization scale $\mu=m_Z$ (solid line),
$\mu=m_Z/2$ (dotted line), and $\mu=2m_Z$ (dash-dotted
line).
\end{minipage}}
\end{picture}
\vskip 2.7cm
\noindent (Initial state photon radiation and 
electroweak corrections are not included.) 
The QCD corrections to the LO result are quite sizable,
as is known also in the massless case. At small values of $y_{cut}$,
 for instance below $y_{cut}\sim 0.01$ for the JADE
algorithm,  perturbation theory is
not applicable. The NLO corrections reduce the dependence on the 
renormalization scale
significinatly, as compared to the leading oder result for the three-jet cross section.    
This renormalization scale dependence, which is also shown in
Figs. 1a,b,  is modest in the whole $y_{cut}$ range exhibited for the Durham
and above $y_{cut}\sim 0.01$ for the JADE algorithm.
\par
The effects of the non-zero $b$ quark mass on the three-jet rate at the $Z$ 
peak is of the order of a few percent, depending on the jet algorithm and on $y_{cut}$.
An interesting application is the determination of the $\overline{\rm MS}$
mass parameter $m_b$ at a high scale by measuring a suitable double ratio of
three-jet fractions, as proposed in \cite{BRS}, computed to NLO in \cite{Val},
and experimentally pursued by the DELPHI collaboration \cite{Fuster}.
In \cite{Ac} we have computed a double ratio that is somewhat differently
defined than the one used in \cite{BRS,Val}. Effects due to the non-zero $b$
quark
mass may also be studied by means of differential two-jet distributions, for instance in the Durham
scheme.

\unitlength 1cm
\begin{picture}(7.5,5)
\put(0,-1.2){\psfig{figure=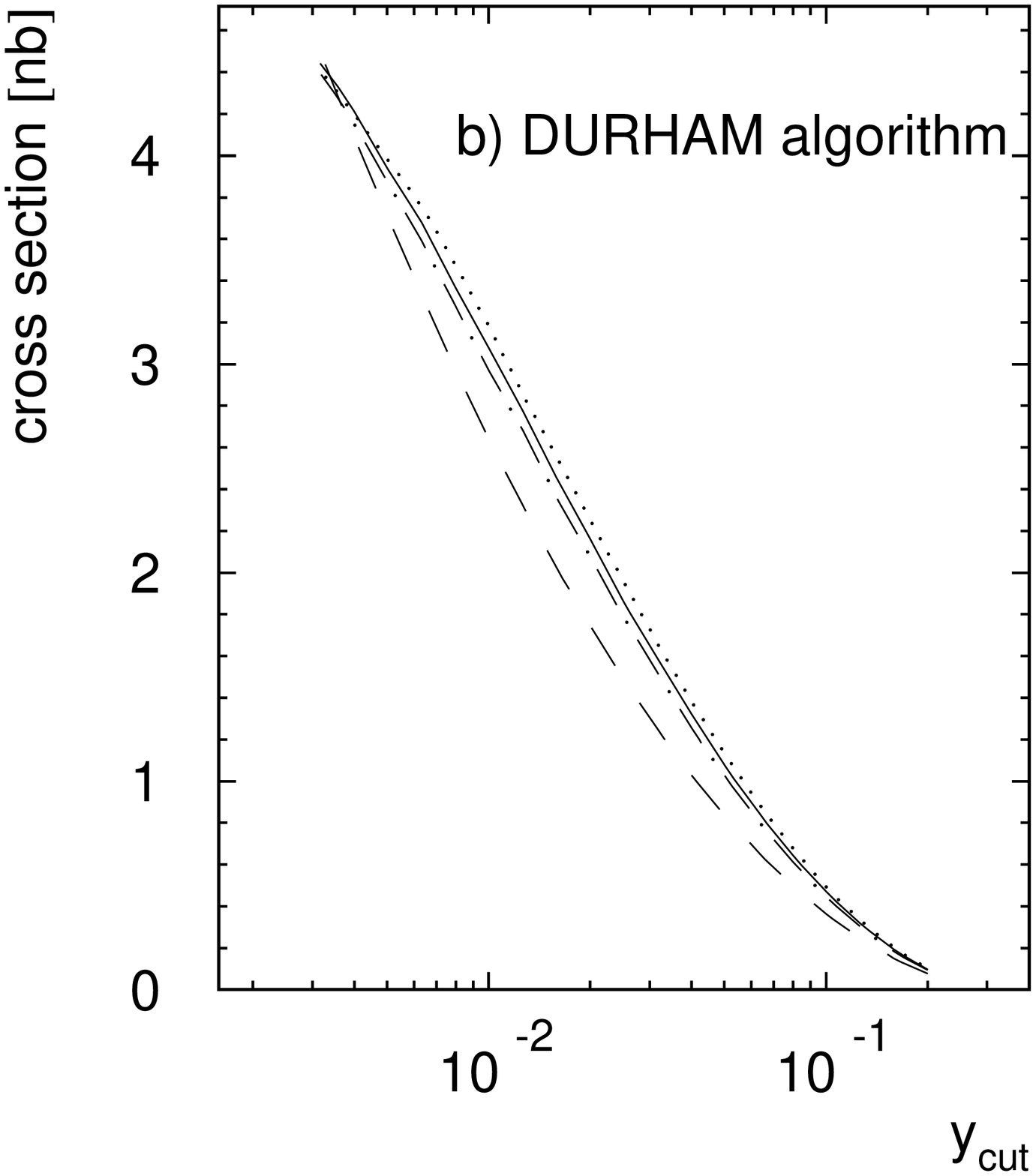,height=6.05cm,width=5.5cm}}
\end{picture}
\par
\vskip 2.7cm

\section{CONCLUSIONS}

We have computed the 
complete differential distributions for $e^+e^-$ annihilation into three and four partons, including the
full quark mass dependence,
 both on and off the $Z$ resonance.  Apart from $\sigma_{NLO}^{3,b}$ these results allow the computation
of a number of other observables at order $\alpha_s^2$. 
Our results may be applied to $b$ and $c$ quark jet production at various
c.m. energies, and
to theoretical investigations 
of top quark production
at very high-energetic electron positron collisions.

\end{document}